# Deep Learning-Based Modulation Detection for NOMA Systems

Wenwu Xie, Jian Xiao, Jinxia Yang, Xin Peng, Chao Yu, Peng Zhu

Hunan Institute of Science and Technology, Hunan, 414006 China

**Abstract**

Since the signal with strong power should be demodulated first for successive interference cancellation (SIC) demodulation in non-orthogonal multiple access (NOMA) systems, the base station (BS) should inform the near user terminal (UT), which has allocated higher power, of modulation mode of the far user terminal. To avoid unnecessary signaling overhead in this process, a blind detection algorithm of NOMA signal modulation mode is designed in this paper. Taking the joint constellation density diagrams of NOMA signal as the detection features, deep residual network is built for classification, so as to detect the modulation mode of NOMA signal. In view of the fact that the joint constellation diagrams are easily polluted by high intensity noise and lose their real distribution pattern, the wavelet denoising method is adopted to improve the quality of constellations. The simulation results represent that the proposed algorithm can achieve satisfactory detection accuracy in NOMA systems. In addition, the factors affecting the recognition performance are also verified and analyzed.

## 1. Introduction

The communication of massive Internet of Things (IoT) requires wireless networks with higher spectral efficiency, lower latency and larger transmission capacity. In the face of the above requirements for higher communication quality, a new multiple access multiplexing method, namely non-orthogonal multiple access (NOMA) was proposed [1]. The research object of this paper is the power domain NOMA, which is the NOMA protocol commonly used at present [2]. In NOMA systems, the base station (BS) exploits the power domain by allocating the same communication resource but different power level to multiple-user (MU) for downlink transmissions. In the downlink NOMA, users with poor channel conditions will be allocated larger power to compensate its low channel gain, which are called far UT, and near UT with better channel conditions will be allocated lower power, which is closer the BS than the far UT. Although interference information is introduced in NOMA system, successive interference cancellation (SIC) technology can be utilized at user terminal (UT) for removing it [3], and thus higher spectral efficiency can be achieved. The signals received by SIC receiver are mixed signals of MU. From the perspective of modulation mode, the signals transmitted by the BS may have different modulation modes. Due to the protocol of NOMA technology, SIC receiver needs to first demodulate the signal desired to far UT, which requires the knowledge of modulation mode for that signal. The general solution is to inform the UT through signaling which can lead to a higher transmission delay in massive IoT scenarios containing enormous devices. Therefore, the implementation of blind modulation detection at near user can reduce signaling overhead for SIC demodulation and further improve the quality of service in NOMA systems.

The NOMA signal is essentially a time-frequency overlapped modulation signal, in which case, the previous single signal modulation recognition algorithm is often no longer applicable. Some research has been done for the modulation recognition of overlapped signals in orthogonal multiple access (OMA) systems, such as using cyclo-stationary theory to extract the feature of signal component [4]. In [5], the maximum likelihood algorithm is used to implement the modulation detection in NOMA systems, which is extended by the ML algorithm in OMA systems [6]. However, the ML algorithm often has a high computational complexity. The work of [7] studied the detection of interference modulation order in downlink NOMA systems, which extracts feature based on Anderson-Darling and then classify by machine learning.

Artificial intelligence technology provides new ideas for designing the next generation of wireless communication systems, which has become a research hotspot in the industry [8, 9]. A deep learning (DL)-aided NOMA system is designed by using long short-term memory network, which can detect the channel characteristics intelligently [10]. In [11], the deep neural network is used to construct the precoder and SIC decoder in MIMO-NOMA system. Both precoding and SIC decoding of the MIMO-NOMA system are jointly optimized, which enables the received signal to be accurately decoded. The application of DL into signal recognition, especially on modulation classification, has attracted most research interests due to its strong feature learning ability [12-14]. Reference [12] surveys the performance gain of general deep neural network architectures for modulation classification is similar when the input data of network is baseband signals. The key to the improvement of recognition performance is to find more discriminative representations of modulated signals for neural network. The constellation diagrams and spectrogram images are used as the input features of convolutional neural networks (CNN) in [13, 14].

In this paper, the joint constellation density diagrams of NOMA signals are selected as the shallow representation, and then a deep residual network [15] for feature learning and classification is designed for realizing the detection of modulation mode of far UT's signal in NOMA system. The main contributions of this paper include:

• Introducing deep learning algorithm to detect modulation mode of NOMA signals.

• Designing preprocessing algorithm to improve the classification performance, which includes wavelet denoising and density extraction.

• Analyzing three factors that affect the detection performance of NOMA modulation signal by numerical experiments.

## 2. System Model

A downlink NOMA scenario is considered in this paper, where a BS communicates with $N$ UTs. $N_c$ UTs share the same sub-channel after user grouping. According the rules of superposition coding [2], the NOMA signal is generated by the superposition of $N_c$ UTs signals with specified power ratios, which is can be expressed as:

$$s = \sum_{i=1}^{N_c} \sqrt{\alpha_i P_t} x^{(i)}, \quad N_c \leq N \tag{1}$$

where $P_t$ is total transmitting power, $\alpha_i$ represents the power radio assigned to $UT_i$ and satisfied $\sum_{i}^{N} \alpha_i = 1$.

The traditional power allocation algorithms can be divided into optimal and suboptimal allocation algorithms [16]. Among them, since the optimal power distribution algorithms, such as the iterative water-filling algorithm, often has too much computation, fractional transmit power allocation algorithms is adopted in this system model. The transmission power of UT $j$ is expressed as:

$$\beta(j) = \frac{1}{\sum_{k \in S} \left( g(k)/n(k) \right)^{-\alpha_{ftpc}}} \left( \frac{g(j)}{n(j)} \right)^{-\alpha_{ftpc}} \tag{2}$$

where $g(j)$ represents channel gain vector. $n(j)$ is the power of noise and interference signals. $S$ is a set of UTs. $\alpha_{ftpc}$ denotes the decay factor, which satisfies $0 < \alpha_{ftpc} \leq 1$. As $\alpha_{ftpc}$ increases, more power is allocated to the Far UT since its channel condition is poor.

For the convenience of analyzing, the scenario considered is that there are two UTs in the same sub-channel is shown in Fig. 1, and the designed algorithm in this paper is still valid for multiple UTs.

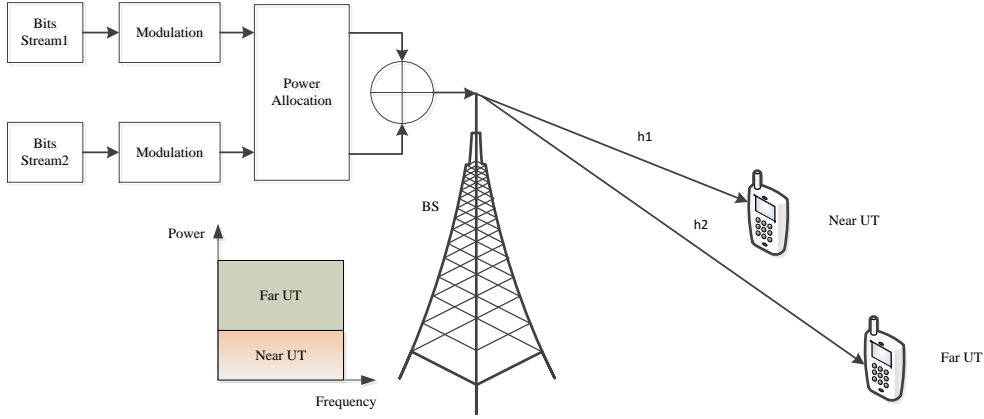

**Fig. 1** A simple downlink two-user NOMA

The received signal of near UT can be written as:
$$s(i) = h(i)(\sqrt{\alpha_1 P_t} x_1(i) + \sqrt{\alpha_2 P_t} x_2(i)) + n(i), \qquad i = 1, 2, ..., N \tag{3}$$

where $x_1(i)$, $x_2(i)$ is modulated signal sent to near user and far user respectively. $\alpha_1$, $\alpha_2$ denotes power radios, which satisfies $\alpha_1 + \alpha_2 = 1$ and $\alpha_2 > \alpha_1$. $P_t$ is total transmitting power. $h(i)$ represent channel gain, which is Rayleigh fading, and $n(i)$ is white Gaussian noise. $N$ is the length of bit stream.

In the decoding process of superimposed coding scheme, the received signal of far UT is decoded directly because $x_1(i)$ are regarded as noise. The received signal of near UT first decodes $x_2(i)$, then reconstructs $x_2(i)$ and cancellation it from the received signal, and then decodes $x_1(i)$. For the above decoding scheme, it is necessary to require the near UTs to know the modulation mode of the far UTs, so that the interference of the far UTs can be removal validly. Therefore, a deep learning algorithm is introduced to realize the modulation blind detection for NOMA system in this paper.

## 3. Proposed Algorithm

### 3.1 The joint constellation density diagrams

The constellation of NOMA signal is the superposition of the constellation of multiple users, which is called the joint constellation diagram. When the component signals of NOMA signal are modulated in different ways or different power allocation ratio, the distribution of its joint constellation will change. So the joint constellation can be used as a feature of NOMA modulation detection, which can be shown in Fig. 2.

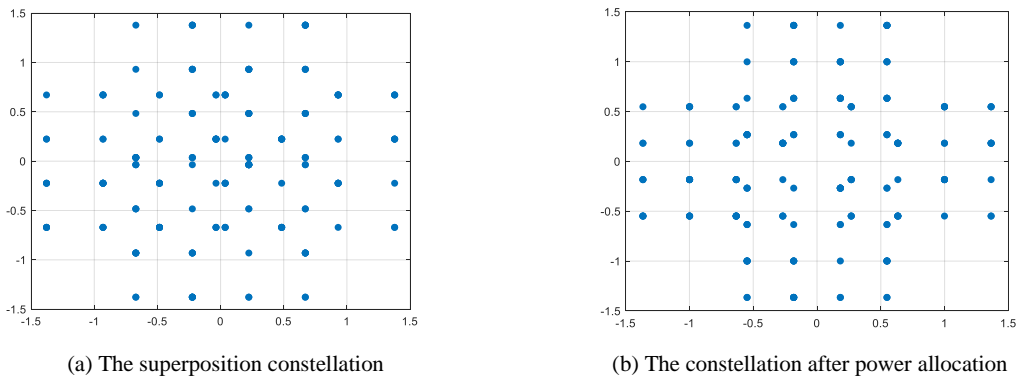

(a) The superposition constellation  (b) The constellation after power allocation

**Fig. 2** The joint constellation diagrams generated from QAM16 signals and $\pi/2$-BPSK signals

When the signal to noise ratio (SNR) is not ideal, the distribution pattern of the constellations will be easily lost, so the quality of the joint constellation diagrams should be improved by denoising first. The traditional denoising method is to filter out the frequency part of the noise by passing the band-pass filter. When the frequency band of the noise is very wide, it is difficult to separate the noise from the effective signal. The multiresolution of wavelet transform (WT) can make the active components of non-stationary signal and noise show different characteristics respectively [17]. The amplitude of wavelet coefficients generated by the WT of the effective signal is larger than the wavelet coefficients of the noise. The WT of signal $x(t)$ can be written as

$$WT_x(a,b) = \frac{1}{\sqrt{a}} \int x(t) \psi^* \left( \frac{t-b}{a} \right) dt \tag{4}$$

where $b$ is time shifting and $a$ represent scale factor. $\psi(t)$ is denoted basic wavelet, * represents the conjugate.

The NOMA signal was separated into real part and imaginary part and computed the wavelet decomposition respectively. Then soft thresholding is applied to the detail wavelet coefficients and wavelet reconstruction is realized according to the original approximation coefficients and the modified detail coefficients. Lastly, the denoising results of the real and imaginary parts are combined to achieve the overall denoising of NOMA signal. The detailed process of wavelet denoising for NOMA signal is shown in Fig. 3.

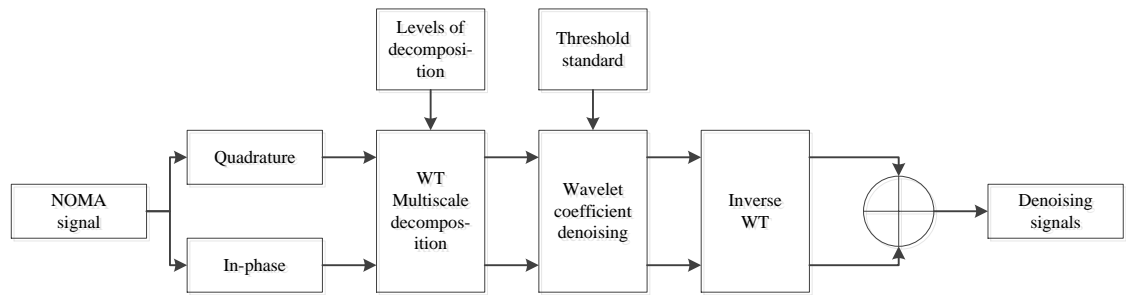

**Fig. 3** The process of wavelet denoising for NOMA signal

Fig. 4 shows the joint constellation diagram after denoising. Simulation parameters are listed in Table I. The NOMA signal can suppress noise and retain effective signal by wavelet denoising, so the joint constellation diagram has more obvious characteristics and tends to an ideal distribution.

**TABLE I** Simulation parameters of preprocessing

| Parameter | Value | Parameter | value |
|---|---|---|---|
| Near UT modulation mode | QAM16 | Basic wavelet | sym8 |
| Far UT modulation mode | π/2-BPSK | Thresholding rule | heursure |
| Near UT SNR | 16 dB | Thresholding type | Soft thresholding |
| Far UT SNR | 10 dB | The level of the WT | 2 |

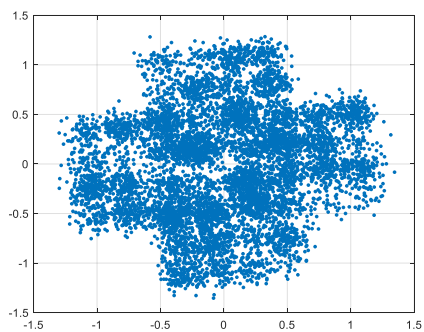
(a) the constellation with noise

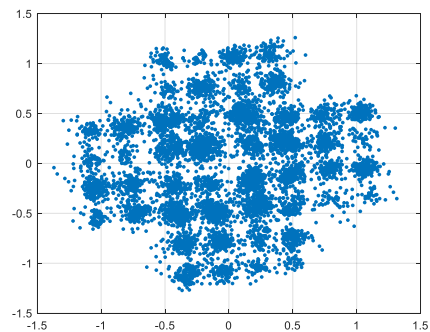
(b) the constellation after wavelet denoising

**Fig. 4** The wavelet denoising of NOMA signal

Considering there may be multiple signal sample points within the range of one pixel, so not every point in a constellation has its value for signal classification. After density extraction of constellation, the pixel value of constellation points can be used as a distinguishing feature. The constellation density is calculated by counting the number of sample points in the fixed region. Then normalizes the constellation density and converts it into the pixel value of the region. Finally, a grayscale image is generated. The density extraction process can be described by Alg. 1 and the joint constellation density diagram is shown in Fig. 5.

| **Algorithm1.** The joint constellation density diagram |
|---|
| Input: The NOMA signal after wavelet denoising $s(i)$ |
| Output: The joint constellation density diagram matrix $G$ with size of $N \times N$ |
| 1: Into two orthogonal groups: <br> $x = imag(s_i), \; y = real(s_i)$ <br> 2: Finding the minimum of $x$ and $y$ <br> $minx = min(x), \; miny = min(y)$ <br> 3: Matrix coordinate transformation <br> $x = ceil(x - min(x)), \; y = ceil(y - min(y))$ <br> 4: Generating the joint constellation matrix $C$ <br>   for $i$ in $range(length(x))$    do <br>     for $j$ in $range(length(y))$    do <br>       $C(x, y) = 1$ <br>     end for <br>   end for <br> 5: Determining statistical area <br> $W = fix(x / N), \; H = fix(y / N)$ <br> 6: Count the number of constellation points in an area and generating the statistical matrix CT <br> 7: Normalization to grayscale matrix G <br> $G(i, j) = \dfrac{CT(i, j) - min(CT)}{(CT) - min(CT)}$ |

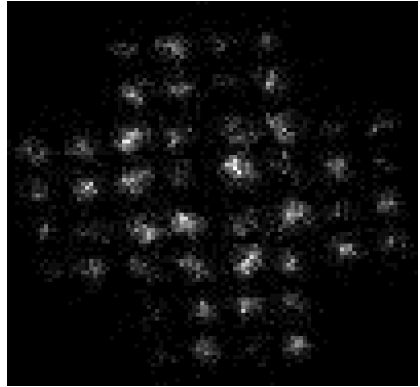

**Fig. 5** the grayscale joint constellation density diagram

The joint constellation density diagram obtained by preprocessing algorithm not only strengthened the differentiation for different NOMA signal constellations, but also adjusted the size of the matrix to make it more suitable as an input matrix for neural network.

3.2 Deep Residual Network

When the neural network have reached a certain depth，which may have reached the optimal network structure, and then still increase the number of network layers, the classification accuracy of neural network will often become worse which is called network degradation. Reference [15] designed a residual learning framework to ease the training of networks. With the introduction of residual learning, neural network is easier for the redundancy layer to realize identity mapping, so as to solve the problem of network degradation.

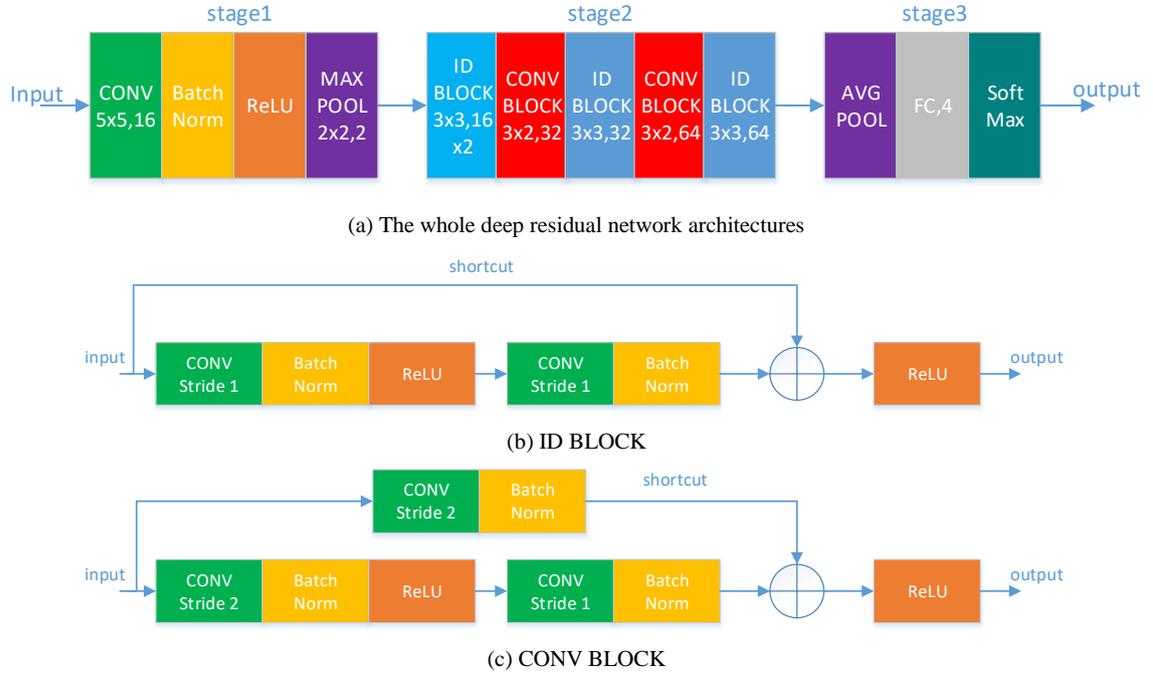

Fig. 6 Deep residual network for modulation detection

In this paper, deep residual network is designed to detect the modulation mode of NOMA signal. The input of the network is the joint constellation density diagrams with size of $100 \times 100$ matrix. The whole network architectures can be divided into three stages, which is shown as Fig .6(a).

In the first stage, the network extracts the characteristics of the bottom layer through a baseline convolution layer. Batch Normalization layer is added after convolution, which pulls the data distribution back to the standard normal distribution with a mean of 0 and a variance of 1, so that the input value of the nonlinear transformation ReLU function falls into the region that is more sensitive to the input, so as to avoid the gradient disappearance problem. Then feature maps are downed sample by max-pooling.

In the second stage, the deep features are excavated through 6 residual blocks. There are two types of residual blocks in Fig. 6(b), (c). One is the identity block in the case of consistent input and output, which is denoted by ID BLOCK. The other is the convolutional block in the case of inconsistent input and output, which is denoted by CONV BLOCK. It includes the convolution operation in the shortcut, and the output matrix is half the size of the input matrix for each CONV BLOCK.

In the last stage, the feature matrixes are downed sample by average-pooling first. After flattening, it passes through a fully connected layer, which contains four output neural nodes. The last layer calculates the softmax with cross-entropy loss of output, which is shown in (5), to determine the results of classification.

$$J = -\sum_i y_i \ln \frac{e^{a_i}}{\sum_{k=1}^{T} e^{a_k}} \qquad (5)$$

where $T$ are all categories of NOMA modulation scheme. $y_i$, $a_i$ represents the i-th output of the network and expected output respectively.

## 4. Simulation and Comparison

In the simulation, each user can choose the modulation mode supported by the current 5G standard, which are $\pi/2$-BPSK, QPSK, QAM16 and QAM64. With the same SNR, there are 1000 NOMA signal samples, and the dividing ratio of training set, validation set and test set is 6:2:2. The SNR of near UT varies from -10 dB to 20 dB with an interval of 2dB. Since the distribution of the joint constellation diagrams are mainly affected by the types of modulation mode of NOMA signal, the number of UTs and power allocation between MU, this paper conducts simulation to verify the influence of these two factors for the modulation detection accuracy of NOMA signal by numerical experiments.

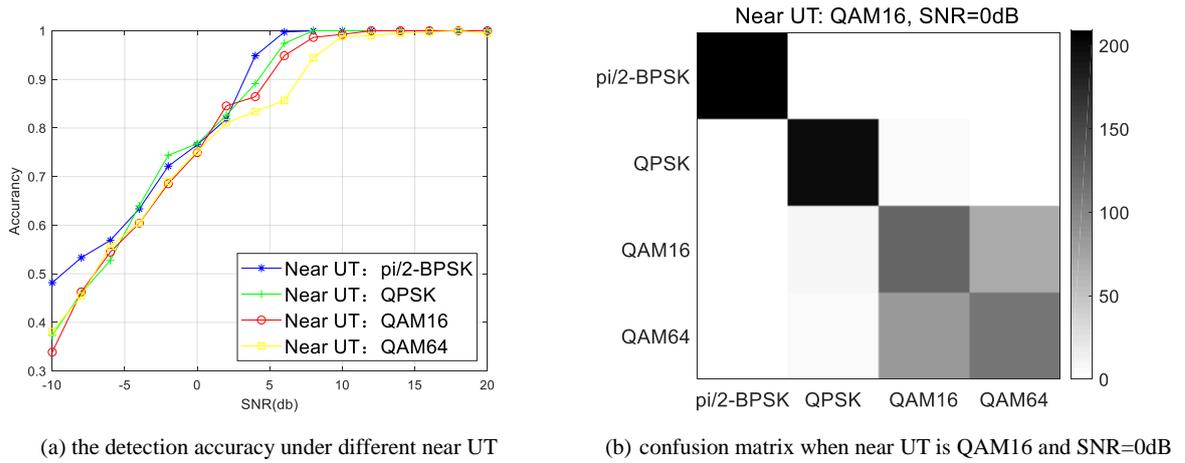

(a) the detection accuracy under different near UT  (b) confusion matrix when near UT is QAM16 and SNR=0dB

**Fig. 7** In the case of fixed power ratio, the detection accuracy under different near UT

When the SNR is above 2dB, Fig. 7(a) represents the modulation recognition accuracy will decrease with the increase of the order of the near UT's modulation signal. Fig. 7(b) shows that when the near UT is the same, the higher order modulation of far UT will misclassify each other. Because the distribution of higher-order constellations is denser, the joint constellation diagrams of NOMA signal are more complex, which are more difficult to distinguish for the neural network. It is worth mentioning that although the constellation of $\pi/2$-BPSK is also of fourth order, its distribution is more characteristic than QPSK. However, under 2dB, the joint constellation diagrams have been polluted by noise to a large extent, and the effect of the wavelet denoising is not obvious. So the recognition rate is roughly the same when the near UT uses the high-order modulation.

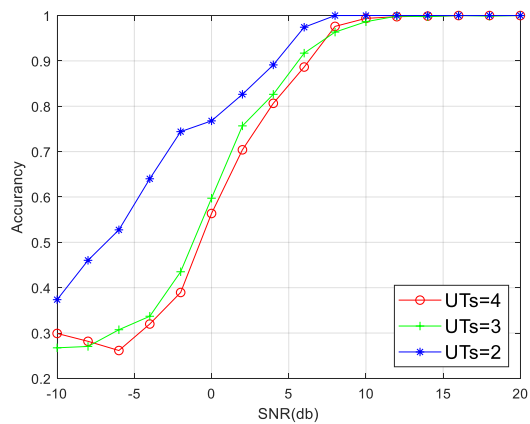

**Fig. 8** The detection accuracy in different numbers of UTs.

As the number of users sharing the same sub-channel increases, the complexity and BER will increase of the SIC decoding scheme in the UTs side, so the situation of different number of users $N_c$ after user grouping was considered in Fig. 8. In the simulation, the number of far UT is fixed to 1, and the number of near UTs is set to 1, 2 and 3, respectively. Different near UTs get different power radios due to different locations and are modulated in the same QPSK scheme. The modulation mode of Far UT could be $\pi/2$-BPSK, QPSK, QAM16 and QAM64. The simulation result shows the detection accuracy will decrease with the decrease of the number of users because the joint constellation becomes more complicated with the increase of users. When the number of UTs is set to 4 and the SNR is above 2dB, the modulation detection results are in the state of random selection.

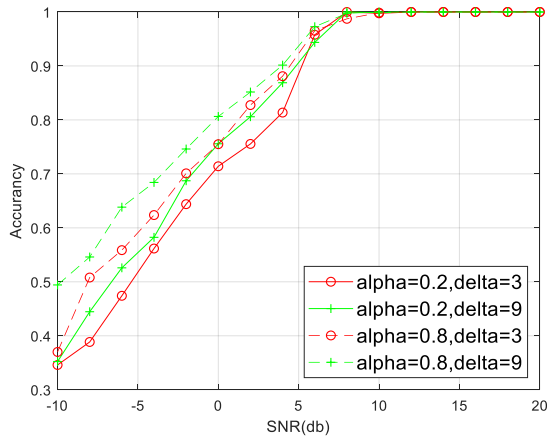

**Fig. 9** In the case of the same near UT, the detection accuracy with different power distribution ratio

When power allocation is carried out in NOMA system, the allocation ratio is positively correlated with the SNR difference between near UT and far UT, which is denoted by 'Delta' in Fig. 9, and the decay factor in fractional power allocation algorithm, which is denoted by 'alpha' in Fig. 9. When the 'alpha' or the 'Delta' is large, the detection accuracy of modulation mode will be improved. This is because the larger the 'alpha' or 'Delta' made the far UT get more large power factor, then the overall distribution of the joint constellation would be biased towards the constellation of far UT, as shown in Fig. 2(a),(b), which will be easy to implement the correct classification.

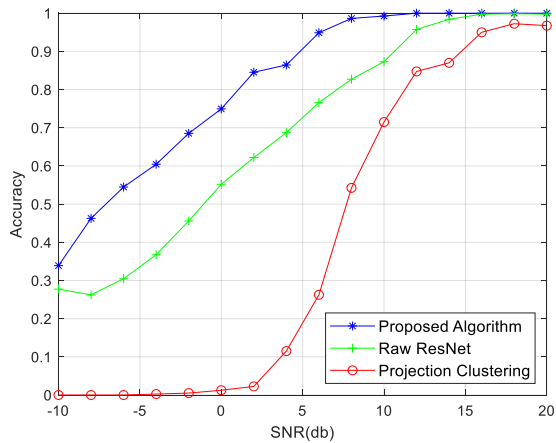

**Fig. 10** The detection accuracy in different modulation recognition algorithms

The proposed algorithm is compared with the traditional modulation recognition algorithm based on projection subtractive clustering of constellations, which is denoted by Projection Clustering in Fig. 10, and using the joint constellations without preprocessing as the input image of the network, which is denoted by Raw ResNet in Fig. 10. The projection clustering algorithm needs to calculate the number of clustering centers first. With the increase of constellation points, the complexity of the clustering algorithm in the traversal process will be higher. When the SNR is not ideal, the representation ability of the joint constellation without wavelet denoising is obviously reduced. So the modulation detection algorithm based on the joint constellation density diagrams and deep residual network is superior to the other two methods in NOMA systems.

## 5. Conclusion

Wireless communication channel has been considered by many researchers is the largest black box in the physical layer. NOMA signals often have a certain extent of distortion and loss at the receiving terminal. The application of machine learning algorithm to help the communication system processes the unknown information over the air, which can achieve greater performance than traditional methods. In this paper, The NOMA signal modulation detection algorithm based on the joint constellation density diagrams and deep residual network is proposed, which will reduce the signaling overhead in the communication system and improve the demodulation efficiency of SIC. However, for the existing high-order modulation mode in 5G, such as 256QAM, its constellation will be more complex, and the classification difficulty will be significantly improved, which is where further research is needed.